\documentclass{ws-ijbc}
\usepackage{ws-rotating}     
\begin{document}

\catchline{}{}{}{}{} 

\markboth{R. Suresh et al.}{Dynamic environment coupling induced synchronized states in coupled time-delayed electronic circuits}

\title{DYNAMIC ENVIRONMENT COUPLING INDUCED SYNCHRONIZED STATES IN COUPLED TIME-DELAYED ELECTRONIC CIRCUITS}

\author{R.~Suresh$^{1}$, K.~Srinivasan$^1$, D.~V.~Senthilkumar$^{3}$, K.~Murali$^4$ M.~ Lakshmanan$^{1,}$\footnote{Author for correspondence}, and J.~Kurths$^{2,3,5}$}

\address{$^1$Centre for Nonlinear Dynamics, School of Physics, Bharathidasan University,\\ Tiruchirapalli - 620 024, India\\
$^{*}$lakshman@cnld.bdu.ac.in}
\address{$^2$Potsdam  Institute for Climate Impact Research, Telegraphenberg, Potsdam D-14473, Germany}
\address{$^3$Institute of Physics, Humboldt University, Berlin D-12489, Germany}
\address{$^4$Department of Physics, Anna University, Chennai, India}
\address{$^5$Institute for Complex Systems and Mathematical Biology, University of Aberdeen, Aberdeen AB24 3FX, United Kingdom}
\maketitle

\begin{history}
\received{(to be inserted by publisher)}
\end{history}

%
\begin{abstract}
We experimentally demonstrate the effect of dynamic environment coupling in a system of coupled piecewise linear time-delay electronic circuits with mutual and subsystem coupling configurations. Time-delay systems are essentially infinite-dimensional systems with complex phase-space properties. Dynamic environmental coupling with mutual coupling configuration has been recently theoretically  shown to induce complete (CS) and inverse synchronizations (IS)~\cite{resmi10} in low-dimensional dynamical systems described by ordinary differential equations (ODEs), for which no experimental confirmation exists. In this paper, we investigate the effect of dynamic environment for the first time in mutual as well as subsystem coupling configurations in coupled time-delay differential equations theoretically and experimentally. Depending upon the coupling strength and the nature of feedback, we observe a transition from asynchronization to CS via phase synchronization and from asynchronization to IS via inverse-phase synchronization in both coupling configurations. The results are corroborated by snapshots of the time evolution, phase projection plots and localized sets as observed from the oscilloscope. Further, the synchronization  are also confirmed numerically from the largest Lyapunov exponents, Correlation of Probability of Recurrence and Correlation Coefficient of the coupled time-delay system. We also present a linear stability analysis and obtain conditions for different synchronized states.
\end{abstract}


\section{\label{sec:level1}Introduction}
Synchronization is a ubiquitous phenomenon often observed in coupled chaotic and hyperchaotic systems \cite{pikovsky01,lakshmanan10}. Depending upon the strength and the nature of coupling, various types of synchronization such as phase synchronization (PS), complete synchronization (CS), generalized synchronization, lag and anticipatory synchronizations, inverse phase synchronization (IPS) and inverse synchronization (IS) have been observed in coupled dynamical systems. All these types of synchronization are achieved in ensembles of dynamical systems through some common coupling schemes, namely linear and nonlinear error feedback couplings \cite{senthil07,senthil06}, coupling via dissimilar and/or time-delayed variables \cite{karnatak07,murphy10,senthil05}, inhibitory coupling \cite{senthil09}, coupling via dynamical relaying \cite{fischer06}, adaptive coupling \cite{ren07} and also systems driven by a common noise \cite{zhou02}, etc.

In many real world systems, synchronization can occur due to interaction through a common dynamic medium and hence the systems also evolve similar to their environment under the influence of the latter. For example, let us consider the synchronization of self-excited nonlinear oscillators suspended on a movable elastic structure. In this case the oscillators interact with each other through the common elastic beam (oscillators are excited by the vibrations of the structure). It has been shown that for given conditions of the elastic structure initially uncorrelated oscillations of each of the oscillators can synchronize with the frequency of the oscillations of elastic beam (both inphase and antiphase synchronizations have been identified) \cite{czolczynski07,kapitaniak13}. Other examples include synchronization of chemical and genetic oscillators \cite{toth06,kuznetsov04,wang05}, synchronized behavior with self pulsating periodic and chaotic oscillations produced by an ensemble of cold atoms interacting with a coherent electromagnetic field \cite{javaloyes08}, synchronization of cells and in coupled circadian oscillators due to common global neurotransmitter oscillation \cite{gonze05}. In all the cases, the coupling function has a dynamics modulated by the system dynamics. In this connection, Resmi et al~\cite{resmi10} have recently shown and verified numerically the existence of various types of synchronization in low-dimensional chaotic systems which are coupled through a dynamic environment without intrinsic or coupling time-delay.

Time-delay systems are an important class of dynamical systems which occur in many real world systems and synchronizing such systems is having potential applications in diverse areas of science, engineering and technology \cite{atay10,rosenblum96,rangan06,sinha05,chen01,buscarino11,pham12} due to their hyperchaotic nature. During the past few years, researchers are interested synchronizing such time-delay systems and mostly all known synchronizations and their transitions are identified and reported in two coupled time-delay systems using numerical simulations \cite{lakshmanan10}. In contrast, investigations on realizing synchronization in time-delay systems from an experimental point of view is very limited compared to their enormous theoretical results. In particular, despite the possibility of realizing time-delay systems using electronic circuits with some effort, experimental realization of most of the promising theoretical results remain largely unexplored using electronic circuits. Experimentally generating and synchronizing such hyperchaotic electronic signals (with multiple positive Lyapunov exponents) is very important where these signals may be used to hide secret messages in the area of secure communication and cryptography \cite{peng03}. Further experimental realizations of theoretical concepts using electronic circuits with time-delay are being used as test beds before implementing them in photonics systems such as liquid state machines to achieve high speed processing, which are usually very expensive \cite{appeltant11,brunner13}.
\begin{figure}
\centering
\includegraphics[width=0.6\columnwidth]{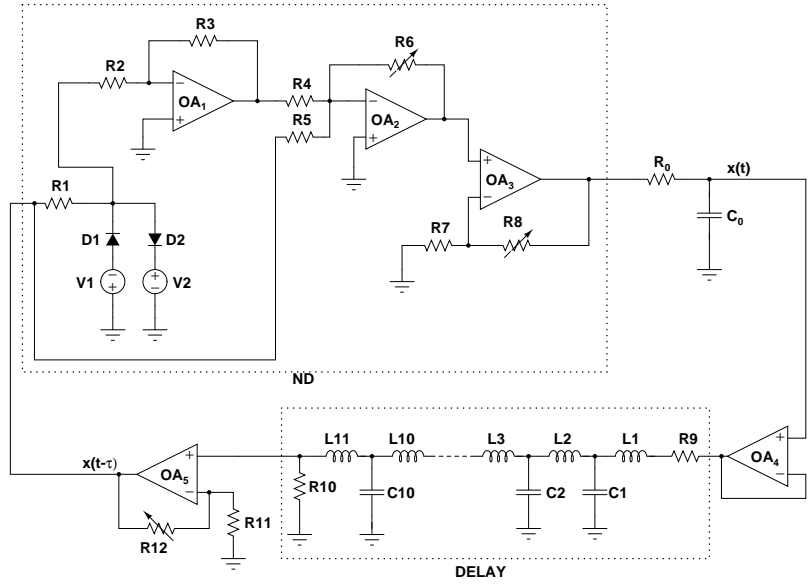}
\caption{\label{fig1} Circuit diagram of a single time delayed feedback oscillator with a nonlinear device (ND) unit, a time-delay unit (DELAY) and a low pass first-order $R_{0}C_{0}$ filter.}
\end{figure}
\begin{figure}
\centering
\includegraphics[width=0.6\columnwidth]{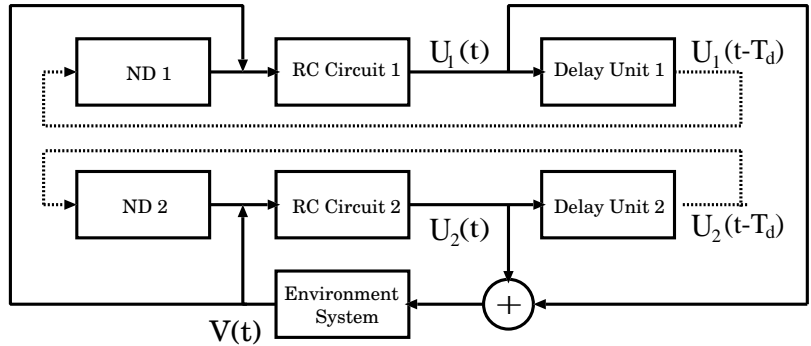}
\caption{\label{fig2} Circuit block diagram of the coupled time delayed feedback oscillator for the mutual coupling configuration (\ref{eqn3}).}
\end{figure}

In this paper, we demonstrate experimentally the occurrence of various types of synchronization and their transitions in indirectly coupled time-delayed electronic circuits with intrinsic time-delay using dynamic environment coupling. We carry out these studies in two different coupling configurations, namely mutual coupling configuration where both the circuits and the environment are mutually sharing their feedback, and subsystem coupling configuration where both circuits are sharing their feedback with the environment, while only one of the circuits is receiving feedback from the environment. It is interesting to note that here synchronization occurs between the independently evolving system and the environmentally affected system. Obviously this has important implications in actual physical situations and to our knowledge this problem has not been studied earlier. Depending upon the coupling strength and the nature of the feedback, we observe different types of synchronization transitions in the coupled circuits which include transition from nonsynchronization to CS via PS and from nonsynchronization to IS via IPS in both coupling configurations. Snapshots of time evolution, phase projection and localized sets of the circuits as observed from the oscilloscope confirm the existence of different synchronized states experimentally along with corresponding numerical results. Further, the transition to different synchronized states can be numerically quantified from the changes in the largest Lyapunov Exponents (LEs), Correlation of Probability of Recurrence (CPR) and Correlation Coefficient (CC) of the coupled systems as a function of the coupling strength. We also present a detailed linear stability analysis and obtain synchronization conditions for different synchronized states in the present system.

The remaining paper is organized as follows: In Sec.~\ref{sec:level2}, we will describe briefly the system employed to demonstrate different synchronized states and explain the circuit realization. In Sec.~\ref{sec:level3} we explain the mutual coupling configuration and present experimental and numerical results to confirm various types of synchronization along with the necessary linear stability analysis. In Sec.~\ref{sec:level4}, we describe the subsystem coupling configuration and demonstrate the occurrence of different synchronization transitions and, finally, we summarize our results with conclusion in Sec.~\ref{sec:level5}.
\begin{figure*}
\centering
\includegraphics[width=1.0\columnwidth]{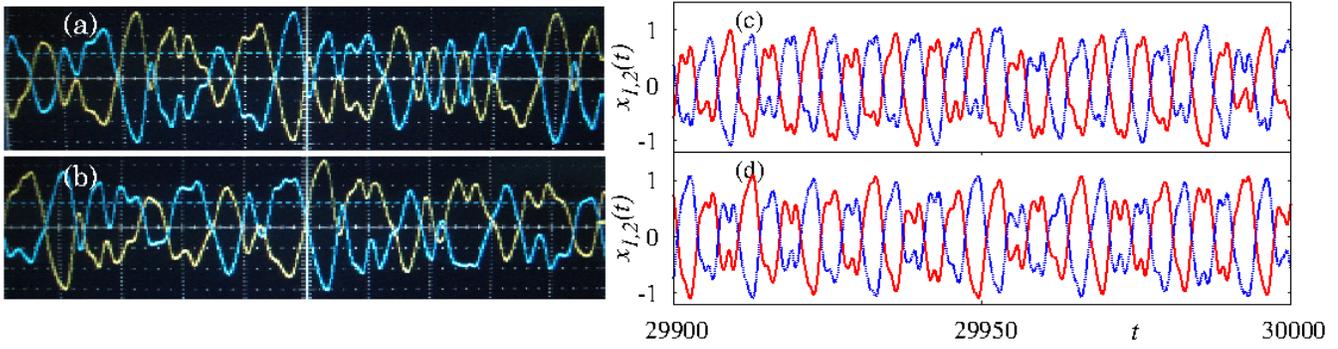}
\caption{\label{fig3} (Color online) Snap shots of the time evolution of the coupled systems (yellow $U_{1}(t)$ and green $U_{2}(t)$) with mutual coupling configuration (\ref{eqn3}) obtained from the oscilloscope indicating the existence of (a) IPS and (b) IS in time-delayed electronic circuits. Vertical scale $2.0v/div$ and horizontal scale $2.0ms/div$. Corresponding numerically obtained time series with ($\beta_{1},\beta_{2})=(1,1)$: (c) IPS for $\varepsilon = 0.6$ and (d) IS for $\varepsilon = 1.5$.}
\end{figure*}
\begin{figure}
\centering
\includegraphics[width=0.7\columnwidth]{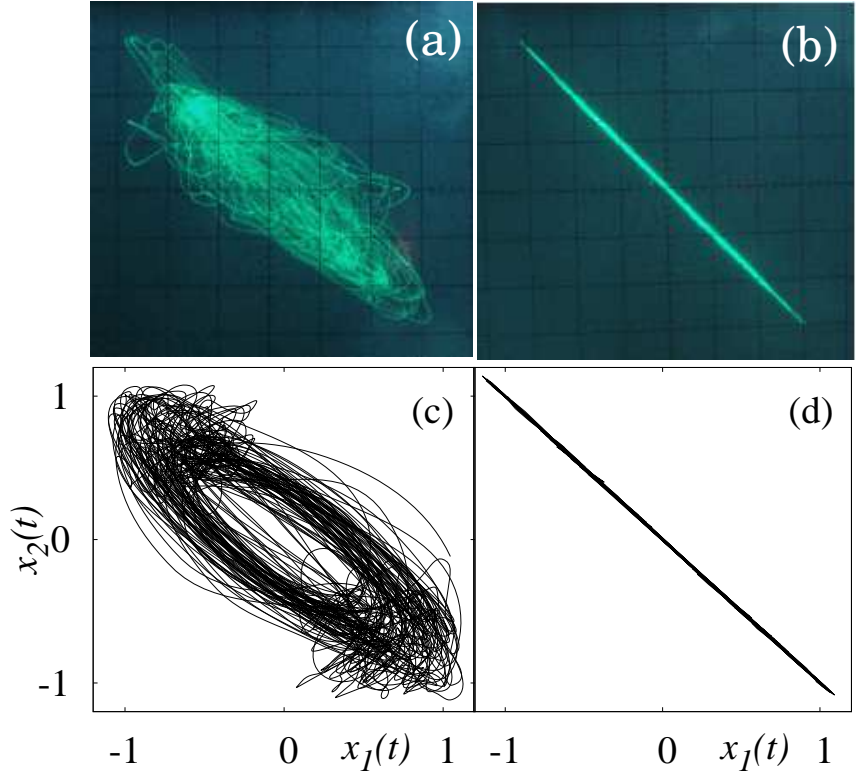}
\caption{\label{fig5} (Color online) Experimental (vertical scale $2.0v/div$ and horizontal scale $2.0v/div$) and numerical observations of phase projections of the coupled systems for mutual coupling configuration: (a) Experimental IPS, (b) experimental IS, (c) numerical IPS for $\varepsilon = 0.6$ and (d) numerical IS for $\varepsilon = 1.5$. }
\end{figure}
\section{\label{sec:level2}Dynamic Environment Coupling}
\subsection{\label{sec:level2ab}System description}
First we consider a single scalar delay differential equation given as
\begin{equation}
\dot{x} = -\alpha x(t)+\beta f(x(t-\tau))
\label{eqn0}
\end{equation}
where the nonlinear function $f(x)$, is represented by a piecewise linear function
\begin{equation}
f(x) = AF^{*}-Bx.
\label{eqn1}
\end{equation}
Here
\begin{eqnarray}
F^{*}=
\left\{
\begin{array}{cc}
-x^{*},&  x < -x^{*}  \\
            x,&  -x^{*} \leq x \leq x^{*} \\
            x^{*},&  x > x^{*}. \\
         \end{array} \right.
\label{eqn2}
\end{eqnarray}
The system parameters are fixed as $\alpha=1.0$, $\beta=1.2$, time-delay $\tau=6.0$, $A=5.2$, $B=3.5$ and $x^{*}=0.7$ throughout the manuscript. For these chosen parameter values the single system (\ref{eqn0}) exhibits a hyperchaotic attractor with three positive LEs \cite{srinivasan11a,srinivasan11b}.
\begin{figure*}
\centering
\includegraphics[width=1.0\columnwidth]{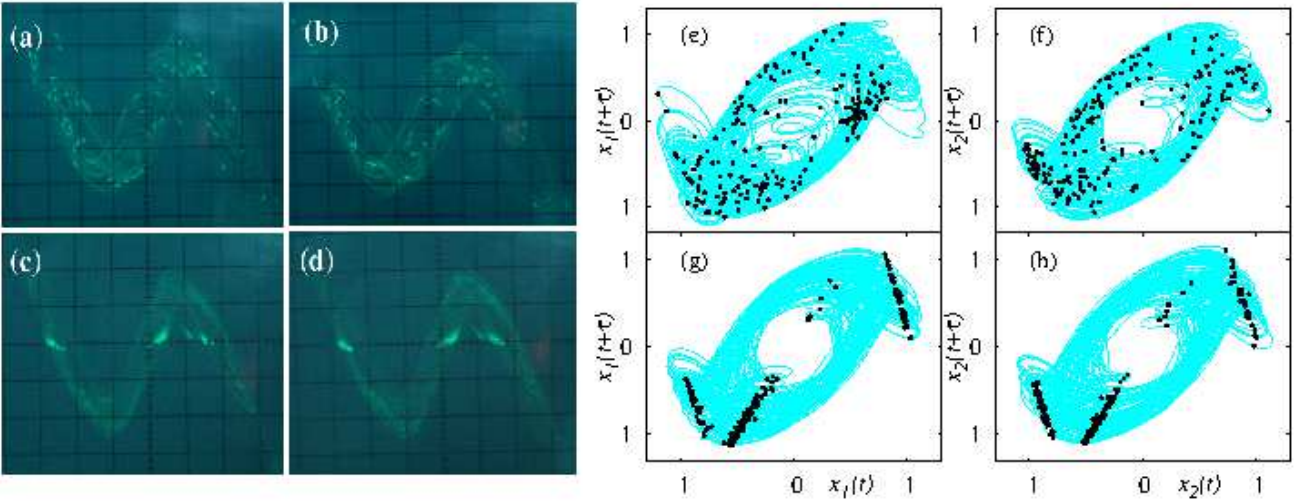}
\caption{\label{fig6} (Color online) (a)-(d) Experimental realization of phase synchronization in (\ref{eqn3}) using the framework of localized sets. (a), (b) The sets are spread over the attractors indicating the absence of phase coherence in the absence of the coupling. (c), (d) For a sufficiently large value of coupling strength, the sets are localized on the attractors which indicates IPS for mutual coupling configuration. Vertical scale $2.0v/div$ and horizontal scale $0.5v/div$. (e)-(h) Corresponding numerically obtained localized sets figures  with ($\beta_{1},\beta_{2})=(1,1)$ in Eq.~(\ref{eqn4}): (e), (f) For $\varepsilon=0$ and (g), (h) For $\varepsilon=0.6$.}
\end{figure*}
\begin{figure}
\centering
\includegraphics[width=0.6\columnwidth]{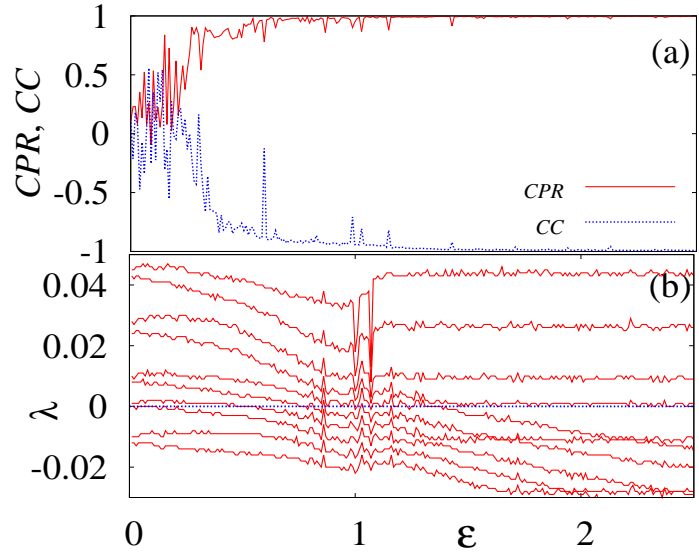}
\caption{\label{fig8} (Color online) (a) CPR (continuous line), CC (dotted line) and (b) spectrum of maximal LEs for ($\beta_{1},\beta_{2})=(1,1)$ in mutual coupling configuration as a function of coupling strength $\varepsilon \in(0,2.5)$.}
\end{figure}
\subsection{\label{sec:level2a}Circuit realization}
The design of the electronic circuit which describes the dynamics of Eq.(\ref{eqn0}) along with the threshold piecewise linear function $f(x)$ is given in Fig.~\ref{fig1}. This circuit consists of a diode based nonlinear device (ND) with two amplification stages ($OA_{1}$ and $OA_{2}$), a time-delay unit (DELAY) and a low-pass first order $R_{0}C_{0}$ filter. Here $\mu A741s$ are engaged as operational amplifiers. $V_{1}$ and $V_{2}$ are the constant voltage sources of all active devices ($\pm 12V$). Using thses voltage values $V_{1}$ and $V_{2}$, one can easily adjust the threshold values of the three segment piecewise function (Eq.~(\ref{eqn2})). By applying the Kirchhoff's laws to this circuit the state equation can be written as $R_{0}C_{0}\frac{dU(t)}{dt}=-U(t)+F[k_{f}(U(t-T_{d}))]$, where $U(t)$ is the voltage across the capacitor $C_{0}$, $U(t-T_{d})$ is the voltage across the delay unit, $T_{d}=n\sqrt{LC}$ is the time-delay, $n$ is the number of LC filter units, and $F[k_{f}(U(t-T_{d}))]$ is the static characteristic of the ND unit.
\begin{figure*}
\centering
\includegraphics[width=1.0\columnwidth]{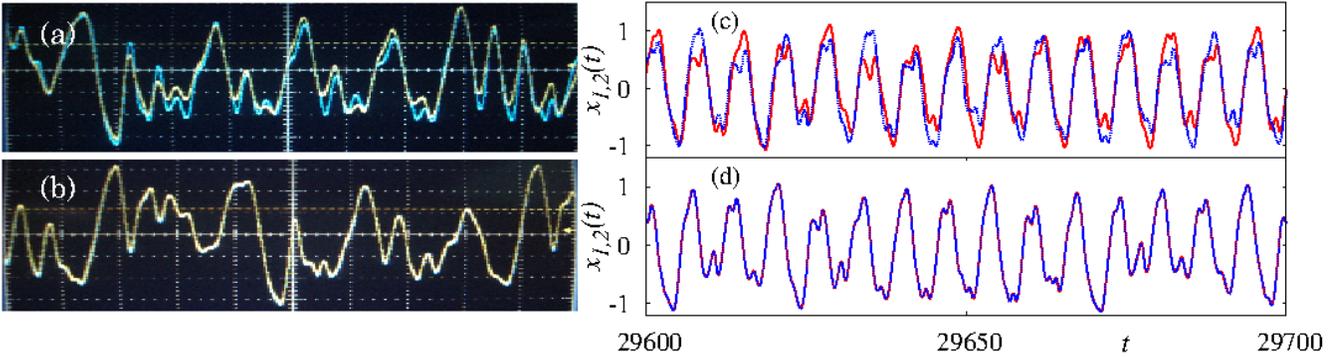}
\caption{\label{fig9} (Color online)(a), (b) Snap shots of the time evolution of both coupled circuits (yellow $U_{1}(t)$ and green $U_{2}(t)$) indicating the existence of (a) PS and (b) CS in coupled time-delayed electronic circuits for mutual coupling configuration. Vertical scale $2.0v/div$ and horizontal scale $1.0ms/div$. Corresponding numerically obtained time series of the coupled systems for ($\beta_{1},\beta_{2})=(1,-1$): (c) PS for $\varepsilon=0.6$ and (d) CS for $\varepsilon=1.5$.}
\end{figure*}
\begin{figure}
\centering
\includegraphics[width=0.6\columnwidth]{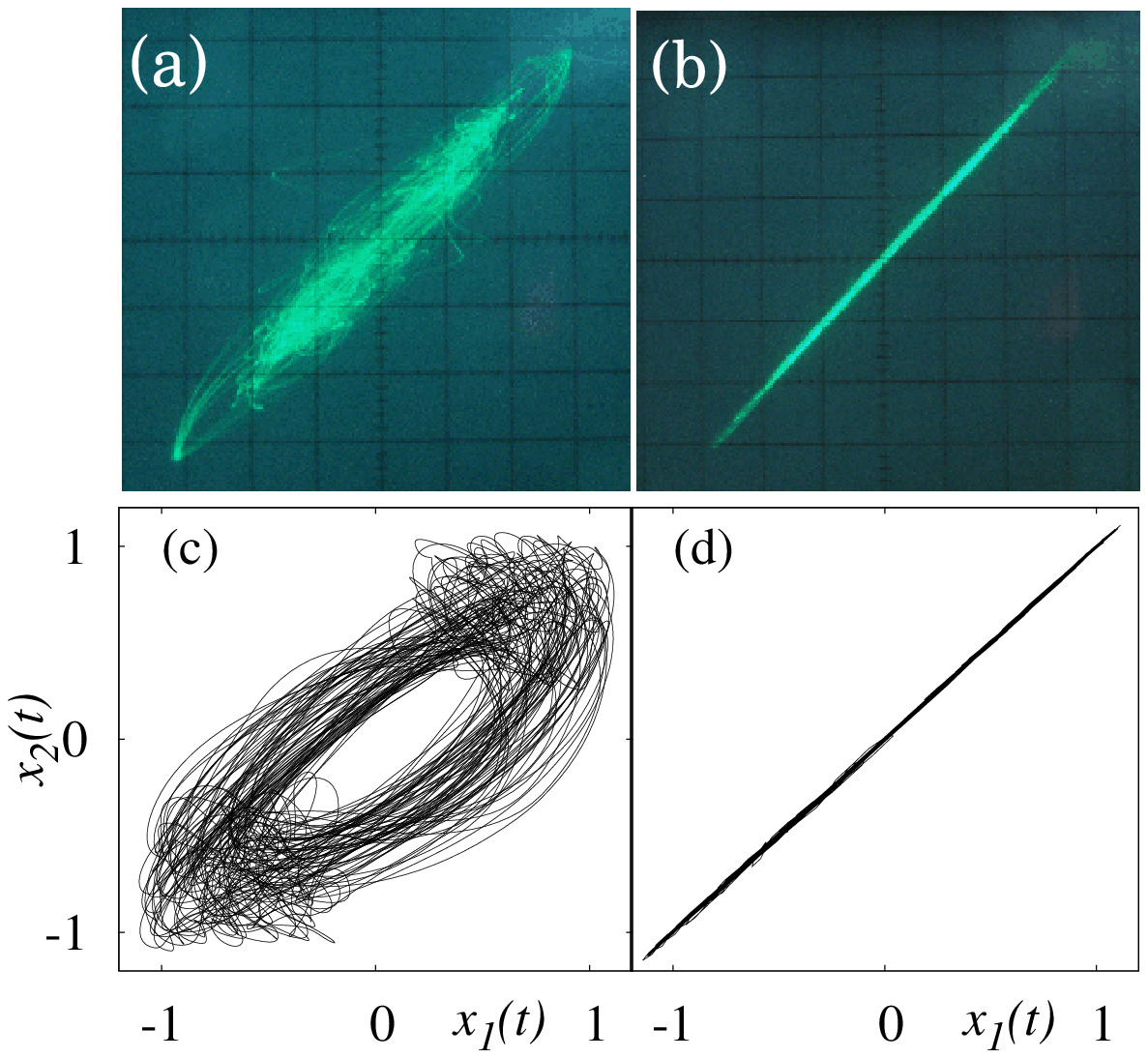}
\caption{\label{fig11} (Color online) Experimental (vertical scale $2.0v/div$ and horizontal scale $2.0v/div$) and numerical phase projection plots of the coupled systems with mutual coupling configuration for ($\beta_{1},\beta_{2})=(1,-1$): (a) Experimental PS, (b) experimental CS, (c) numerical PS for $\varepsilon=0.6$ and (d) numerical CS for $\varepsilon=1.5$. }
\end{figure}

To study the circuit equation, we transform it to the dimensionless form (\ref{eqn0}) by defining the dimensionless variables and parameters as $x(t)=\frac{U(t)}{U_{s}}$, $t^{\prime}=\frac{t}{R_{0}C_{0}}$, $\tau=\frac{T_{d}}{R_{0}C_{0}}$, $k_{f}=\beta$, and $t^{\prime}\rightarrow t$. A nonzero $U_s$ is chosen such that $ND(U_s)=U_s$. The circuit parameters are fixed as $R_{1}=1K\Omega$, $R_{2}=R_{3}=10K\Omega$, $R_{4}=2K\Omega$, $R_{5}=3K\Omega$, $R_{6}=10.4K\Omega$ (trimmer-pot), $R_{7}=1K\Omega$, $R_{8}=5K\Omega$ (trimmer-pot), $(R_{9}=R_{10}=1K\Omega$, $R_{11}=10K\Omega$, $R_{12}=20K\Omega$ (trimmer-pot), $R_{0}=1.86K\Omega$, $C_{0}=100nF$, $L_{i}=12mH$ ($i=1,2,\cdots,11$), $C_{i}=470nF$ ($i=1,2,\cdots,10$), $n=10$. $T_{d}=0.751ms$, $R_{0}C_{0}=0.268ms$, and so the time-delay $\tau\approx2.8$ for the chosen circuit parameter values.

\section{\label{sec:level3}Mutual Coupling Configuration}
We now construct a circuit which consists of a system of two identical time-delayed sub-circuits with a threshold piecewise-linear nonlinearity and are coupled indirectly through a common environment. Here both the circuits and the environment are mutually sharing their feedback with each other and the state equation for the coupled circuit can be written as
\begin{subequations}
\begin{eqnarray}
R_{0}C_{0} \frac{dU_{1}(t)}{dt} &=& -\alpha^{\prime}U_{1}(t)+f[k_{f}U_{1}(t-T_{d})]+ \nonumber \\
\varepsilon^{\prime}_{1} \beta^{\prime}_{1}V(t),\\
R_{0}C_{0} \frac{dU_{2}(t)}{dt} &=& -\alpha^{\prime}U_{2}(t)+f[k_{f}U_{2}(t-T_{d})]+ \nonumber \\
\varepsilon^{\prime}_{1} \beta^{\prime}_{2}V(t),\\
R_{0}C_{0} \frac{dV(t)}{dt} &=& -k^{\prime}V(t)-\frac{\varepsilon^{\prime}_{2}}{2}[\beta^{\prime}_{1}U_{1}(t)+\beta^{\prime}_{2}U_{2}(t)],
\end{eqnarray}
\label{eqn3}
\end{subequations}
where $U_{1}(t)$ and $U_{2}(t)$ correspond to the output variables of each circuit and $V(t)$ is the output of the environmental equation. The schematic diagram for this coupled circuit is sketched in Fig.~\ref{fig2}. By defining the normalized variables and parameters as above and $x_{1,2}(t)=\frac{U_{1,2}(t)}{U_{s}}$, $y(t)=\frac{V(t)}{V_{s}}$ one obtains the equivalent dimensionless equation as follows:
\begin{subequations}
\begin{eqnarray}
\dot{x}_1(t)&=&-\alpha x_1(t)+\beta f[x_{1}(t-\tau)]+\varepsilon_{1}\beta_{1}y,\\
\dot{x}_2(t)&=&-\alpha x_2(t)+\beta f[x_{2}(t-\tau)]+\varepsilon_{1}\beta_{2}y,\\
\dot{y} &=& -k y-\frac{\varepsilon_{2}}{2} (\beta_{1}x_{1}+\beta_{2}x_{2}),
\end{eqnarray}
\label{eqn4}
\end{subequations}
\begin{figure*}
\centering
\includegraphics[width=1.0\columnwidth]{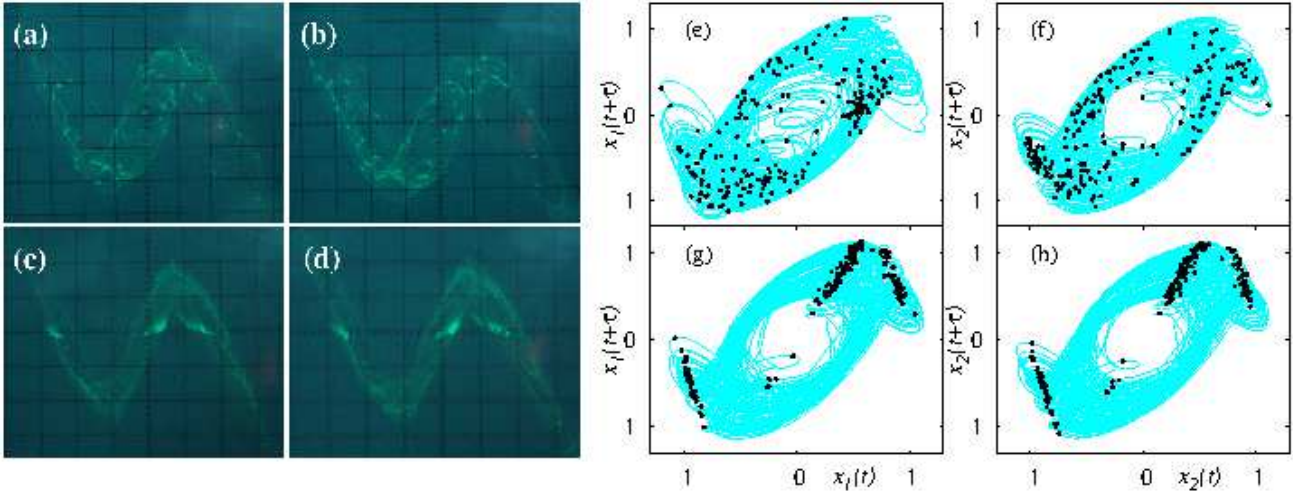}
\caption{\label{fig12} (Color online) (a)-(d) Experimental realization of phase synchronization using the framework of localized sets in the case of mutual coupling configuration. (a), (b) The sets are spread over the attractors indicating the absence of phase coherence. (c), (d) The sets are localized on the attractors which indicates phase synchronization. Vertical scale $2.0v/div$ and horizontal scale $0.5v/div$. (e)-(h) Corresponding numerically obtained localized sets for the case ($\beta_{1},\beta_{2})=(1,-1)$: (e), (f) for $\varepsilon=0$ and (g) ,(h) for $\varepsilon=0.6$.}
\end{figure*}

where, $\alpha^{\prime}=\alpha$, $\beta^{\prime}_{1,2}=\beta_{1,2}$, $\varepsilon^{\prime}_{1,2}=\varepsilon_{1,2}$ and $k^{\prime}=k$. Here the two systems $x_{1}(t)$ and $x_{2}(t)$ are not directly coupled to each other, instead they are coupled through a coupling function ($y$) which has a dynamics modulated by the system dynamics. $\varepsilon_{1}$ is the strength of the feedback to the systems and $\varepsilon_{2}$ is the strength of the feedback to the environment (coupling parameters). $\beta_{1}$ and $\beta_{2}$ are the nature of feedback from and to the environment, respectively. $k$ is the damping parameter and we choose it as $k=1$. In the absence of feedback from the systems to the environment, the strength of the environment decays exponentially fast as $k>0$. Note that in the absence of delay $(\tau=0)$, the system of Eqs.~(\ref{eqn4}) reduces to the dynamical system studied by Resmi et al ~\cite{resmi10}.
\begin{figure}
\centering
\includegraphics[width=0.6\columnwidth]{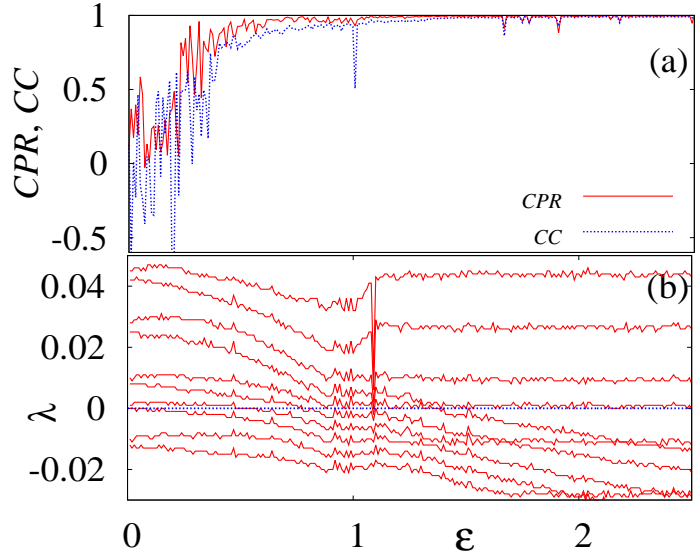}
\caption{\label{fig14} (Color online) (a) CPR (continuous line), CC (dotted line) and (b) spectrum of maximal LEs for $(\beta_{1},\beta_{2})=(1,-1)$ in mutual coupling configuration as a function of coupling strength $\varepsilon \in(0,2.5)$.}
\end{figure}
\subsection{\label{sec:level3a}Experimental and numerical observations}
We have experimentally observed different types of synchronization in the coupled electronic circuits which are also confirmed using numerical simulations. Depending upon the feedback strength $\beta_{1}$ and $\beta_{2}$ we observe two types of synchronization transitions in the coupled systems. When $\beta_{1}$ and $\beta_{2}$ are of the same sign, for example ($\beta_{1},\beta_{2})=(1,1)$, we observe a transition from non-synchronization to IS via IPS and when $\beta_{1}$ and $\beta_{2}$ are of different signs, ($\beta_{1},\beta_{2})=(1,-1)$, then we observe a transition from non-synchronization to CS via PS as a function of the coupling strength.

\subsubsection{The case $\beta_{1}$ and $\beta_{2}$ with same sign}
First, we consider the case ($\beta_{1}, \beta_{2})=(1, 1)$ and for simplicity we have chosen the value of the coupling strengths as $\varepsilon_{1}=\varepsilon_{2}=\varepsilon$. In the absence of the coupling ($\varepsilon=0$) both circuits oscillate independently, and for sufficiently large value of the coupling strength the phase difference between the circuits is exactly $\pi$, that is the systems exhibit IPS. On increasing the coupling strength to further larger values, IS occurrs between the circuits.
\begin{figure}
\centering
\includegraphics[width=0.6\columnwidth]{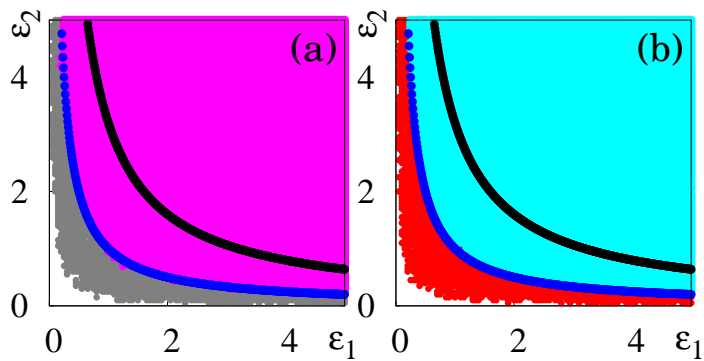}
\caption{\label{fig15} (Color online) (a), (b) Numerically obtained two parameter diagrams in ($\varepsilon_{1}-\varepsilon_{2}$) plane show various types of synchronization states in the case of mutual coupling configuration for ($\beta_{1},\beta_{2})=(1,1$) and ($\beta_{1},\beta_{2})=(1,-1$), respectively. We mark PS and IPS states when CPR $>0.96$. CS and IS states are marked when CC $>0.96$ and CC $>-0.96$, respectively. In both figures, the white color indicates desynchronized state, grey color in Fig.~\ref{fig15}(a) and red color (dark grey) in Fig.~\ref{fig15}(b) represent IPS and PS regimes, respectively. IS state is represented by pink color (dark grey in Fig.~\ref{fig15}(a)) and light blue color (light grey in Fig.~\ref{fig15}(b)) indicate the CS region. The black line indicates the analytically obtained stability condition of the outer regime of the piecewise linear function [Eq.~(\ref{eqn2})] and blue (dark grey) filled circles indicates the stability condition for the middle region of the piecewise linear function.}
\end{figure}

Snapshots of the wave forms of the circuits as seen in the oscilloscope are shown in Fig.~\ref{fig3}. The existence of IPS between the circuits is depicted in Fig.~\ref{fig3}(a) where the circuits are evolving with a phase difference of $\pi$ but still the amplitudes are uncorrelated. Figure~\ref{fig3}(b) shows the realization of IS between both systems where both phase and amplitude are correlated and occur exactly opposite to each other. In the corresponding numerical analysis of Eq.~(\ref{eqn4}), we obtain IPS for $\varepsilon=0.6$ which is depicted in Fig.~\ref{fig3}(c) and this synchronization can also be confirmed both experimentally and numerically using the phase projection plots which are shown in Figs.~\ref{fig5}(a) and \ref{fig5}(c), respectively. On further increase of the coupling strength to $\varepsilon=1.5$, the systems exhibit IS where the maxima of both systems occur exactly opposite to each other for as depicted in Fig.~\ref{fig3}(d). The corresponding experimental and numerical phase projection plots of the systems are given in Figs.~\ref{fig5}(b) and \ref{fig5}(d), respectively.

Phase coherence of the systems is further qualitatively visualized both experimentally and numerically using the framework of localized sets \cite{pereira07}. The basic idea of this characterization is that the set of points obtained by sampling the time-series of the system 1 whenever the maximum occurs in system 2 is plotted along with the attractor of system 1 and vice versa. If the coupled systems are said to be phase synchronized then the sets are localized on the attractor, otherwise they spread over the entire attractor implying asynchronization. This provides an easy and efficient way to detect phase synchronization even in non-phase-coherent and high dimensional attractors.

The experimental realization of the localized sets can be obtained as follows: The maxima of the systems state variables of the circuits 1 and 2 are taken as our reference points to demonstrate localized sets. Using the circuit given in Fig. 7.12 in pp. 147 of \cite{lakshmanan95}, we generate the impulse whenever the input signal of the system 2 attains maximum. While the attractor of the system 1 is in the $X-Y$ channel of the oscilloscope, we feed the impulse signal to the $Z$-input. Whenever the impulse hits the attractor, one can see the bright spot in the attractor of the system 1. The same is observed when the impulse generated from the system 1 and superimposed to the attractor of the system 2. The experimental observation of the localized sets is shown in Fig.~\ref{fig6}. In the absence of the coupling ($\varepsilon=0$), the sets are distributed over the entire attractor which corresponds to the absence of phase coherence. Figures \ref{fig6}(a) and \ref{fig6}(b) show that the attractors of the two systems along with the sets for the case of nonphase synchronization.  The corresponding numerically obtained figures are plotted in Figs.~\ref{fig6}(e) and \ref{fig6}(f) for $\varepsilon=0$. For sufficiently large value of the coupling strength, the sets are localized on their corresponding attractors which confirm a perfect phase locking of the systems [Figs.~\ref{fig6}(c) and \ref{fig6}(d)]. The corresponding numerical figures are plotted in Figs.~\ref{fig6}(g) and \ref{fig6}(h) for the value of $\varepsilon=0.6$.
\begin{figure}
\centering
\includegraphics[width=0.6\columnwidth]{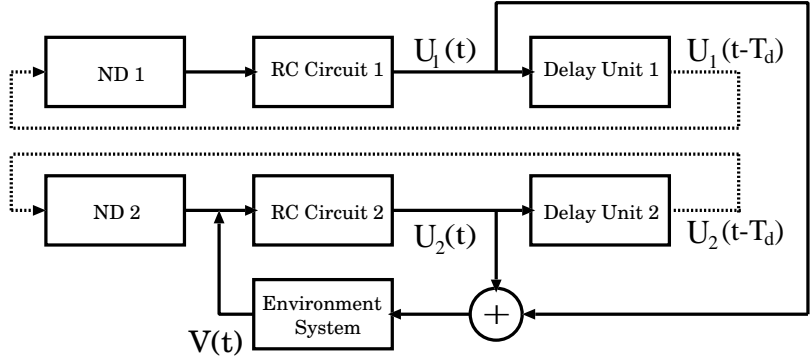}
\caption{\label{fig16} Circuit block diagram of the coupled time delayed feedback oscillator for subsystem coupling configuration (\ref{eqn12}).}
\end{figure}
\begin{figure*}
\centering
\includegraphics[width=1.0\columnwidth]{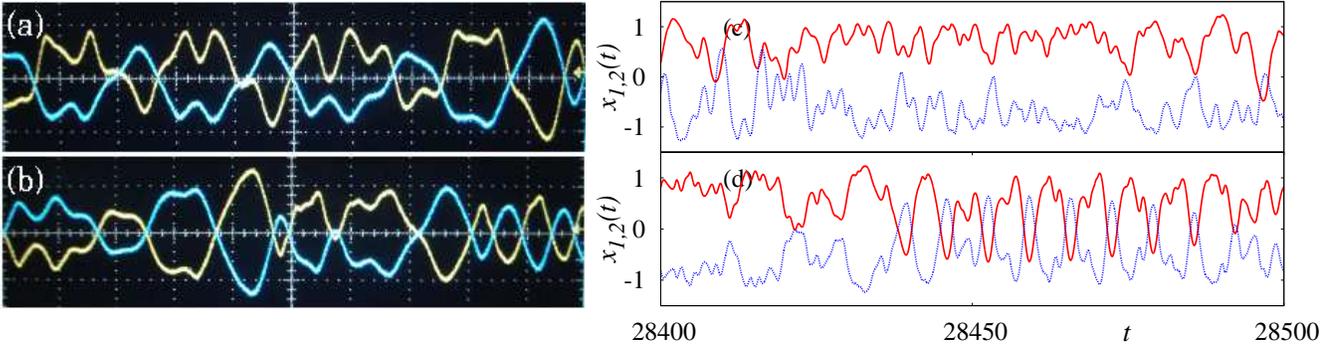}
\caption{\label{fig17} (Color online) Snap shots of the time evolution of both coupled
circuits (yellow $U_{1}(t)$ and green $U_{2}(t)$) indicating the existence of (a) IPS and (b) IS in coupled time-delayed electronic
circuits for subsystem coupling configuration (\ref{eqn11}). Vertical scale $5.0v/div$ and horizontal scale $2.0ms/div$. Corresponding numerically obtained time series of the coupled systems with ($\beta_{1},\beta_{2})=(1,1$): (c) IPS for $\varepsilon=1.2$ and (d) IS for $\varepsilon=2.0$.}
\end{figure*}

Next, the transition from nonsynchronization to IS via IPS can be characterized by changes in the spectrum of maximal LEs. Also the phase coherence is further quantified using the index CPR. Complete and inverse synchronizations can be quantified using the CC. These are given by the expressions,
\begin{eqnarray}
CPR &=&\langle \bar{P_1}(t)\bar{P_2}(t)\rangle/\sigma_1\sigma_2, \label{eqn5a} \\
CC &=& \frac{\langle(x_{1}(t)-\langle x_{1}(t)\rangle)(x_{2}(t)-\langle x_{2}(t)\rangle)\rangle}{\sqrt{\langle(x_{1}(t)-\langle x_{1}(t)\rangle)^{2}\rangle \langle(x_{2}(t)-\langle x_{2}(t)\rangle)^{2}\rangle}},
\label{eqn5b}
\end{eqnarray}
where $\langle\quad\rangle$ brackets indicate time average. Using Eq. (\ref{eqn5a}), we calculate the index CPR. Here $\bar{P}_{1,2}$ means that the mean value has been subtracted, $\sigma_{1,2}$ are the standard deviations of $P_{1}$ and $P_{2}$ and $P(t)$ is a generalized autocorrelation function based on recurrence properties \cite{marwan07}. If the phases of the systems are perfectly locked, then the probability of recurrence is maximal at the time $t$ and $CPR\approx1$, otherwise the maxima do not occur simultaneously and hence one can expect a drift in both the probability of recurrence resulting in low values of CPR. Using Eq. (\ref{eqn5b}), we calculate the CC to characterize the CS and IS between the systems. If both systems are in CS state then, $CC\approx1$ and for IS state $CC$ will be $\approx-1$.
\begin{figure*}
\centering
\includegraphics[width=1.0\columnwidth]{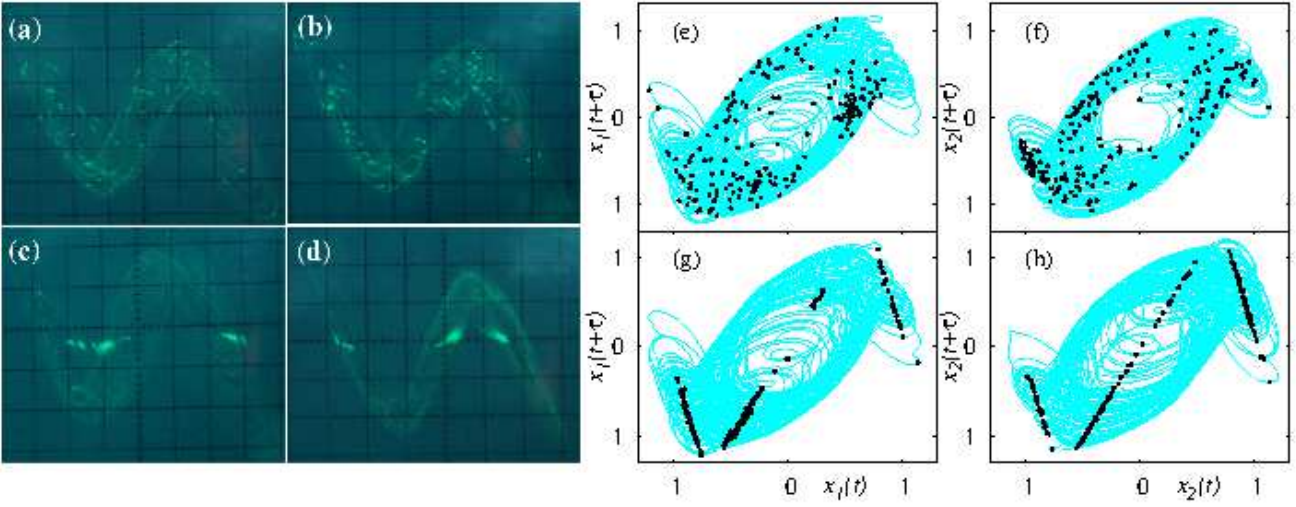}
\caption{\label{fig20} (Color online) (a)-(d) Experimental realization of the framework of localized sets in subsystem coupling configuration with $\beta_{1},\beta_{2}=(1,1)$. (a), (b) The sets are spread over the attractors indicating the absence of phase coherence. (c), (d) The sets are localized on the attractors which indicates the occurrence of IPS. Vertical scale $2.0v/div$ and horizontal scale $0.5v/div$. (e)-(h) Corresponding numerically obtained localized sets figures: (e), (f) For $\varepsilon=0$ and (g), (h) for $\varepsilon=1.2$.}
\end{figure*}

In Fig.~\ref{fig8}(a) we have plotted the numerically calculated CPR (continuous line), CC (dotted line) and in Fig.~\ref{fig8}(b) we have plotted the ten largest LEs of the coupled systems as a function of the coupling strength $\varepsilon\in(0,2.5)$. In the absence of the coupling ($\varepsilon=0$), the index CPR and CC are near to zero which confirms that the systems are evolving without any synchronization with six positive LEs (three for each systems). If the coupling strength increases, the index CPR also starts to increase towards unit value and for $\varepsilon=0.56$, CPR is oscillating near unity (CPR $\approx 0.97$) whereas the value of CC becomes negative which indeed confirms the onset of IPS. Additional confirmation comes from the changes in the LEs, where the zeroth LE of the coupled system becomes negative at $\varepsilon=0.56$ which indicates the existence of IPS. If we increase the coupling strength further, the CC of the coupled systems decreases and reaches the value $\approx -0.99$ at $\varepsilon=1.39$. At this value of $\varepsilon$, except for the three positive LEs, all the other positive LEs of the coupled systems become negative which indeed confirms the existence of IS in the coupled time-delay systems.

\subsubsection{The case $\beta_{1}$ and $\beta_{2}$ with opposite sign}
Next, we consider the case with ($\beta_{1},\beta_{2})=(1,-1)$ (different signs). Now we observe a transition from nonsynchronization to CS via PS as a function of the coupling strength. In this case, for lower values of coupling strength the individual circuits evolve independently, while for $\varepsilon=0.56$ the circuits exhibit PS and for further larger value of $\varepsilon$ ($\varepsilon=1.39$)
both systems attain a CS state.

The experimental snap shots and numerical plots of the time series of the coupled systems are depicted in Figs.~\ref{fig9}(a) and \ref{fig9}(c), respectively, for $\varepsilon=0.6$ indicating that the circuits are evolving with PS. This is also verified (both experimentally and numerically) using the phase projection plots [Figs.~\ref{fig11}(a) and \ref{fig11}(c)], respectively. On further increase of the coupling strength to $\varepsilon= 1.5$, the circuits exhibit CS with each other which is depicted in Figs.~\ref{fig9}(b) (experimetnal) and \ref{fig9}(d) (numerical). The corresponding experimental and numerical phase projection plots of the systems are given in Figs.~\ref{fig11}(b) and \ref{fig11}(d), respectively.

PS is further confirmed by the frame work of localized sets. The experimental observations of localized sets are presented in Fig.~\ref{fig12}. The sets are distributed over the entire attractors which confirm no synchronization in the absence of the coupling [Figs.~\ref{fig12}(a) and \ref{fig12}(b)]. The corresponding numerical figures are shown in Figs.~\ref{fig12}(e) and \ref{fig12}(f) for $\varepsilon=0$. For a sufficiently large value of coupling strength, the sets are localized on the attractors which confirm a perfect phase locking between the systems [Figs.~\ref{fig12}(c) and \ref{fig12}(d)]. The equivalent numerical figures are given in Figs.~\ref{fig12}(g) and \ref{fig12}(h) for the value of coupling strength $\varepsilon=1.5$. We note here that we cannot distinguish IPS and PS using the framework of localized sets. This method is used to characterize the phase coherence/incoherence depending on the localization of the sets on the attractor. To differentiate IPS from PS we use other plots such as snapshot of the time evolution and phase projection plots.

In Fig.~\ref{fig14}(a) we have plotted the index CPR, CC and in Fig.~\ref{fig14} (b) we have plotted ten maximal LEs of the coupled systems as a function of the coupling strength ($\varepsilon \in (0,2.5)$). In the absence of the coupling, the index CPR and CC $\approx 0$ confirming that the systems are independently oscillating without any synchronization. Once the coupling strength increases beyond $0.5$ ($\varepsilon=0.56$), the index CPR increases and reaches a value close to unity ($CPR \approx0.97$) and CC becomes $\approx 0.8$, confirming the onset of PS but still the amplitudes are uncorrelated [Fig.~\ref{fig14}(a)]. PS is also confirmed from the changes in the LEs where the zeroth LE of the coupled system becomes negative at $\varepsilon>0.5$ [Fig.~\ref{fig14}(b)]. Further, for $\varepsilon =1.39$, the CC of the system reaches the unit value ($\approx 0.99$) which is depicted in Fig.~\ref{fig14}(a) and for this $\varepsilon$ value, except for the three positive LEs, all the other positive LEs of the coupled system become negative confirming the existence of CS [Fig.~\ref{fig14}(b)].

To identify the global picture of occurrence of different synchronization transition regimes, we have plotted the phase diagrams (obtained using numerical simulation) in the two parameter plane of coupling strengths ($\varepsilon_{1}, \varepsilon_{2}$). We use the index CPR to mark PS and IPS regimes when CPR $>0.96$. In addition, CS region is marked when CC $>0.96$  and IS region is marked when CC $>-0.96$. The phase diagrams in the ($\varepsilon_{1}-\varepsilon_{2}$) plane for the coupled time-delay systems are shown in Fig.~\ref{fig15}(a) and Fig.~\ref{fig15}(b) for ($\beta_{1},\beta_{2}$) = ($1,1$) and ($1,-1$), respectively. In both figures the white color represents the desynchronized state, grey color in Fig.~\ref{fig15}(a) and red color (dark grey) in Fig.~\ref{fig15}(b) correspond to IPS and PS regimes, respectively. IS is represented by pink color (dark grey) in Fig.~\ref{fig15}(a) and light blue color (light grey) indicates CS regime in Fig.~\ref{fig15}(b).
\begin{figure}
\centering
\includegraphics[width=0.6\columnwidth]{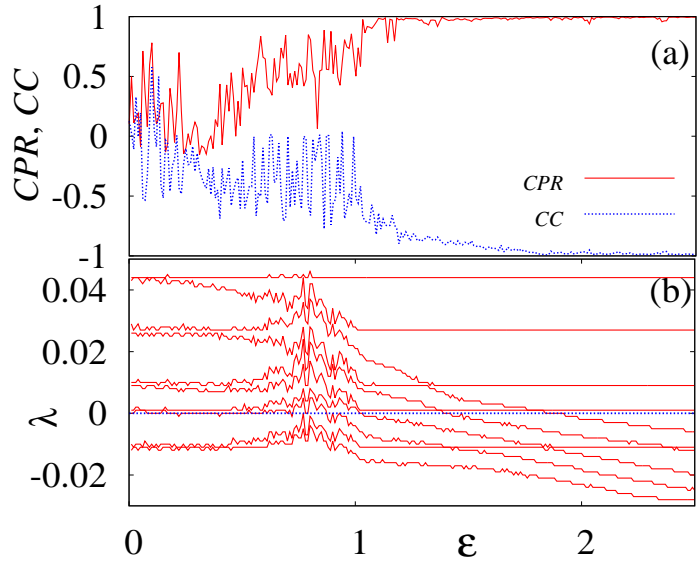}
\caption{\label{fig22} (Color online) (a) CPR (continuous line), CC (dotted line) and (b) spectrum of maximal LEs for $(\beta_{1},\beta_{2})=(1,1)$ in subsystem coupling configuration as a function of coupling strength $\varepsilon \in(0,2.5)$.}
\end{figure}
\subsection{\label{sec:level3b}Linear stability analysis}
In this section, we arrive at the stability condition for the synchronized states of the two dynamically coupled time-delay systems. To find the stability of the synchronization, we apply an infinitesimal perturbation to the system (\ref{eqn4}). Let $\xi_{1}, \xi_{2}$ and $z$ represent the perturbations. Then Eq.~(\ref{eqn4}) can be rewritten as
\begin{subequations}
\begin{eqnarray}
\dot{\xi}_{1}&=&-\alpha \xi_{1}+\beta f^{\prime}(x_{1\tau})\xi_{1\tau}+\varepsilon_{1}\beta_{1}z,\\
\dot{\xi}_{2}&=&-\alpha \xi_{2}+\beta f^{\prime}(x_{2\tau})\xi_{2\tau}+\varepsilon_{1}\beta_{2}z,\\
\dot{z} &=& -kz-\frac{\varepsilon_{2}}{2} [\beta_{1}\xi_{1}+\beta_{2}\xi_{2}].
\end{eqnarray}
\label{eqn6}
\end{subequations}
Here $f^{\prime}(x_{\tau})=f^{\prime}[x(t-\tau)]$ is the derivative of Eq.~(\ref{eqn1}) and for the two outer regimes of Eq.~(\ref{eqn2}), it can be written as $f^{\prime}(x_{\tau})=-B$ and for the middle region $f^{\prime}(x_{\tau})=(A-B)$. It is also to be noted that in the recent papers \cite{resmi10, banerjee13}, the time average of the derivative of the nonlinear function $f(x)$ is approximated as a constant. This makes the entire stability analysis to be generic for all forms of nonlinearity when the systems are coupled through the environmental coupling scheme (\ref{eqn4}). However, the derivative of $f(x_\tau)$ in our analysis naturally becomes different constants in different regions because of the piecewise linear nature of $f(x_{\tau})$ and hence we assume $f^{\prime}(x_{1\tau})=f^{\prime}(x_{2\tau})=\phi$, for generality. Here $\phi=-B$ in the two outer regions and $\phi=(A-B)$ in the middle region.

Equation (\ref{eqn6}) is difficult to solve in the present form and so we consider the special case of complete synchronization state ($x_{1}=x_{2}$). Now Eq.~(\ref{eqn6}) can be simplified by defining
\begin{equation}
\xi_{0} = \beta_{1}\xi_{1}+\beta_{2}\xi_{2}, \quad \chi = \beta_{2}\xi_{1}-\beta_{1}\xi_{2}.
\label{eqn7}
\end{equation}
So Eq.~(\ref{eqn6}) becomes
\begin{subequations}
\begin{eqnarray}
\dot{\xi}_{0}&=&-\alpha \xi_{0}+\beta\phi\xi_{0\tau}+\varepsilon_{1}z(\beta_{1}^{2}+\beta_{2}^{2}),\\
\dot{z} &=& -kz-\frac{\varepsilon_{2}}{2} \xi_{0},\\
\dot{\chi}&=&-\alpha\chi+\beta\phi\chi_{\tau}.
\end{eqnarray}
\label{eqn8}
\end{subequations}
Note that Eqs.~(\ref{eqn8}a) and (\ref{eqn8}b) are coupled while Eq.~(\ref{eqn8}c) is uncoupled from the other two and corresponds to the perturbation of the uncoupled time-delay system (\ref{eqn0}) confirming the underlying dynamics is chaotic/hyperchaotic for chosen parameter values. We note here that for ($\beta_{1},\beta_{2}$)=(1,-1), $\xi_{0}$ is a perturbation to the complete synchronization manifold and for ($\beta_{1},\beta_{2}$)=(1,1) corresponds to a perturbation to the inverse synchronization manifold. The synchronization state determined by $\xi_{0}$ is stable if the real part of all the eigenvalues of Eqs.~(\ref{eqn8}a) and (\ref{eqn8}b) are negative.

For ($\beta_{1},\beta_{2}$)=(1,1) or (1,-1), ($\beta_{1}^{2}+\beta_{2}^{2})=2$ and eliminating $z$ from Eq.~(\ref{eqn8}), we get
\begin{equation}
\ddot{\xi}_{0}+(\alpha+k)\dot{\xi}_{0}-\beta\phi\dot{\xi}_{0\tau}-\beta\phi k \xi_{0\tau}+(\alpha k+\varepsilon_{1}\varepsilon_{2})\xi_{0}=0.
\label{eqn9.1}
\end{equation}
Consider a solution of the form
\begin{equation}
\xi_{0}=A e^{\lambda t},
\end{equation}
which on substitution in (\ref{eqn9.1}) and after simple algebraic manipulation yields the relation
%
%
%
%
%
%
%
\begin{equation}
\varepsilon_{1}\varepsilon_{2}>|k(\beta\phi-\alpha)|. \label{eqn9.7a}
\end{equation}
%
We have considered the middle region of the piecewise linear function where most of the system dynamics is confined. For ($\beta_{1},\beta_{2}$)=(1,-1) the stable synchronization is achieved at the threshold value
\begin{equation}
\varepsilon_{1c}=\frac{|k(\beta (A-B)-\alpha)|}{\varepsilon_{2c}}.
\label{eqn10.1}
\end{equation}
and for the two outer regimes ($|x|>x^{*}$) we obtain the condition
\begin{equation}
\varepsilon_{1c}=\frac{|-k(\beta B+\alpha)|}{\varepsilon_{2c}},
\label{eqn10}
\end{equation}

From Eqs.~(\ref{eqn10}) and (\ref{eqn10.1}) one can obtain the threshold value of the coupling strength $\varepsilon_{1}$ for the synchronized state. For the above chosen values and the feedback ($\beta_{1},\beta_{2})=(1,1)$ we obtain a transition from asynchronization to an IS state as shown in Fig.~\ref{fig15}. The blue (dark grey) filled circles in Fig.~\ref{fig15}(a) corresponds to the stability condition for the middle regime and the black line to the stability condition of the two outer regimes of the piecewise linear function. This shows that the numerically obtained IS regime exactly fits with the analytically obtained stability condition of the middle regime of the piecewise linear function where most of the dynamics is confined for the asymptotic synchronized state. Similar transition curves are also observed for transition from asynchronization to CS in the case of $(\beta_{1},\beta_{2})=(1,-1)$ [Fig.~\ref{fig15}(b)].
\begin{figure*}
\centering
\includegraphics[width=1.0\columnwidth]{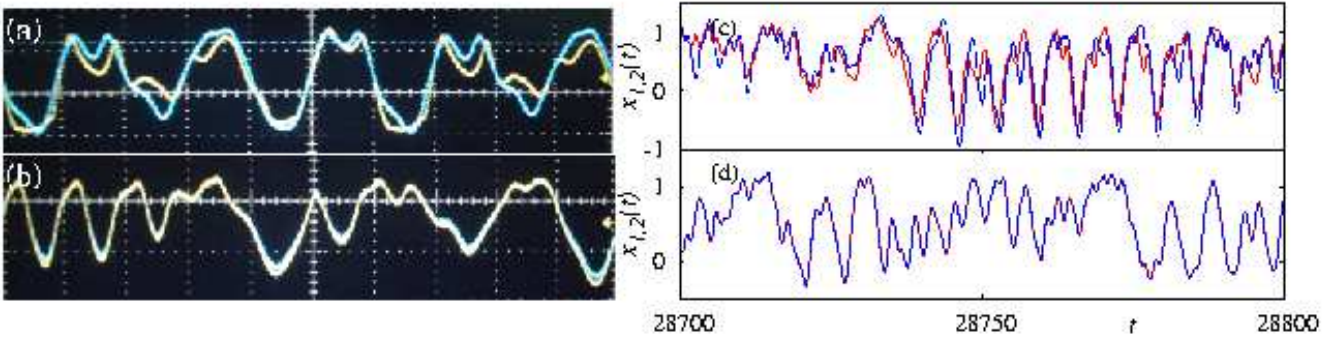}
\caption{\label{fig23} (Color online) (a), (b)Snap shots of the time evolution of both coupled circuits (yellow $U_{1}(t)$ and green $U_{2}(t)$) indicating the existence of (a) PS and (b) CS in coupled time-delayed electronic circuits for subsystem coupling configuration (\ref{eqn11}). Vertical scale $5.0v/div$ and horizontal scale $1.0ms/div$. Corresponding numerically obtained time series with ($\beta_{1},\beta_{2})=(1,-1$): (c) PS for $\varepsilon=1.2$ and (d) CS for $\varepsilon=2.0$}
\end{figure*}
\section{\label{sec:level4}Subsystem Coupling Configuration}
Further, we have also considered a second form of coupling called subsystem coupling configuration, where both circuits share their feedback with the environment, while only one of the circuits is receiving a feedback from the environment. The schematic circuit block diagram for this coupling configuration is sketched in Fig.~\ref{fig16} and the state equations can be given as follows:
\begin{subequations}
\begin{eqnarray}
R_{0}C_{0} \frac{dU_{1}(t)}{dt} &=& -\alpha^{\prime}U_{1}(t)+f[k_{f}U_{1}(t-T_{d})],\\
R_{0}C_{0} \frac{dU_{2}(t)}{dt} &=& -\alpha^{\prime}U_{2}(t)+f[k_{f}U_{2}(t-T_{d})]+ \nonumber\\
\varepsilon^{\prime}_{1} \beta^{\prime}_{2}V(t),\\
R_{0}C_{0} \frac{dV(t)}{dt} &=&-k^{\prime}V(t)-\frac{\varepsilon^{\prime}_{2}}{2}[\beta^{\prime}_{1}U_{1}(t)+\beta^{\prime}_{2}U_{2}(t)],
\end{eqnarray}
\label{eqn11}
\end{subequations}
and the corresponding dimensionless equations for this configuration can be written as
\begin{subequations}
\begin{eqnarray}
\dot{x}_1(t)&=&-\alpha x_1(t)+\beta f[x_{1}(t-\tau)],\\
\dot{x}_2(t)&=&-\alpha x_2(t)+\beta f[x_{2}(t-\tau)]+\varepsilon_{1}\beta_{2}y,\\
\dot{y} &=& -ky-\frac{\varepsilon_{2}}{2} (\beta_{1}x_{1}+\beta_{2}x_{2}).
\end{eqnarray}
\label{eqn12}
\end{subequations}
\begin{figure*}
\centering
\includegraphics[width=1.0\columnwidth]{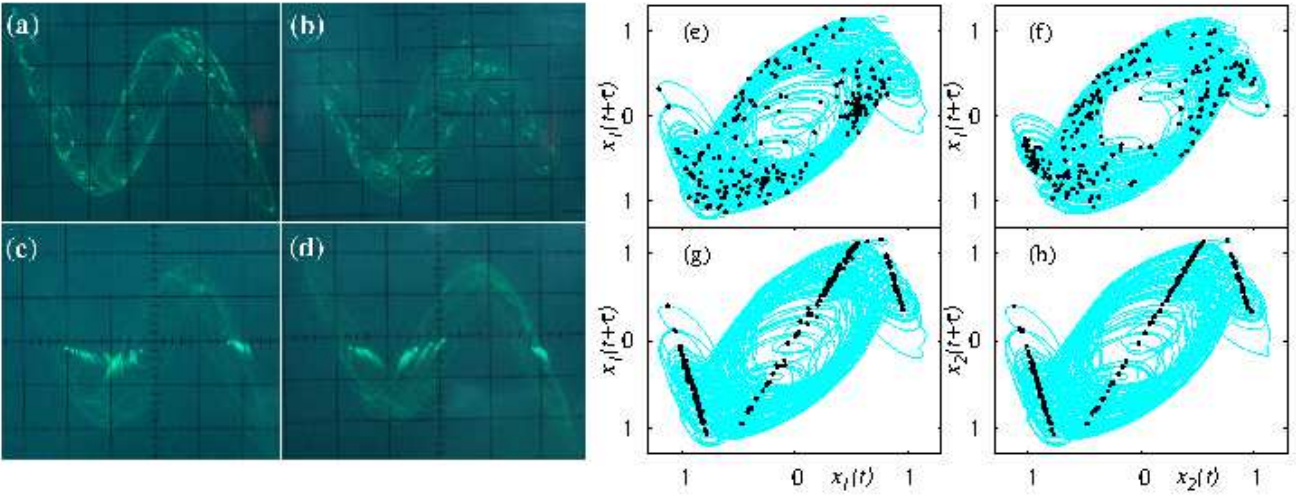}
\caption{\label{fig26} (Color online) (a)-(d) Experimental realization of the phase synchronization using the framework of the localized sets in subsystem coupling configuration. (a), (b) The sets are spread over the attractors indicating the absence of phase coherence. (c), (d) The sets are localized on the attractors which indicates phase synchronization. Vertical scale $2.0v/div$ and horizontal scale $0.5v/div$. (e)-(h) Corresponding numerically obtained localized sets figures for the case ($\beta_{1},\beta_{2})=(1,-1)$: (e), (f) for $\varepsilon=0$ and (g), (h) for $\varepsilon=1.2$.}
\end{figure*}

Here the circuits of $U_{1}(t)$ and $U_{2}(t)$ ($x_{1}(t)=\frac{U_{1}(t)}{U_s}$ and $x_{2}(t)=\frac{U_{2}(t)}{U_s}$) are sharing their feedback with the environment but only the circuit of $U_{2}(t)$ is receiving a feedback from the environment. The parameters are fixed as in the previous case.

\subsection{The case $\beta_{1}$ and $\beta_{2}$ with same sign}
First we consider the case ($\beta_{1},\beta_{2})=(1,1)$ and obtain the same kind of transition from IPS (for $\varepsilon>1.0$) to IS (for $\varepsilon>1.9$). The experimental and numerical snap shots of the time evolution of the coupled systems show IPS for the coupling strength $\varepsilon=1.2$ which is evident from Figs.~\ref{fig17}(a) and \ref{fig17}(c), respectively. If we increase the coupling to larger values, both circuits attain IS at $\varepsilon=2.0$ as depicted in Figs.~\ref{fig17}(b) (experimental ) and \ref{fig17}(d) (numerical).

The localized sets plots are again presented to confirm the existence of IPS. The experimentally obtained attractors along with the sets are shown in Figs.~\ref{fig20}(a) and \ref{fig20}(b) where the sets are distributed over the entire attractors in the absence of any coupling and the corresponding numerical figures are depicted in Figs.~\ref{fig20}(e) and \ref{fig20}(f) for $\varepsilon=0$. For sufficiently large values of the coupling strength, one can observe that the sets are localized on the attractors confirming the perfect locking of the phases of the systems [Figs.~\ref{fig20}(c) and \ref{fig20}(d)]. The equivalent numerical figures are plotted in Figs.~\ref{fig20}(g) and \ref{fig20}(h) for $\varepsilon=1.2$.

The transition from IPS to IS can again be confirmed by plotting the index CPR, CC and the changes in the maximal LEs as a function of the coupling strength. In the absence of the coupling, the systems are evolving without any synchronization (with six positive LEs) and so CPR and CC $\approx 0$. For $\varepsilon=1.02$, the value of index CPR increases towards unity and reaches the value CPR $\approx0.97$, whereas CC becomes negative which confirms the onset of IPS state [Fig.~\ref{fig22}(a)]. This transition is also confirmed from the changes in the LEs where the zeroth LE of the coupled system becomes negative for this value of $\varepsilon$ indicating the existence of IPS which is evident from Fig.~\ref{fig22}(b). On further increase of the coupling strength to $\varepsilon=1.91$ the systems exhibit IS transition where the CC becomes $\approx-0.99$ [Fig.~\ref{fig22}(a)] and from Fig.~\ref{fig22}(b), we notice that except for the three positive LEs, all the other positive LEs become negative confirming the existence of IS in the coupled time-delay system.

\subsection{The case $\beta_{1}$ and $\beta_{2}$ with opposite sign}
Finally, we consider the case ($\beta_{1},\beta_{2})=(1,-1)$. We obtain a transition from no synchronization to CS via PS in the coupled time-delayed electronic circuit. The experimental snap shots of the time evolution of both circuits are displayed in Fig.~\ref{fig23}(a) indicating that the systems exhibit PS and the numerically obtained figure is depicted in Fig.~\ref{fig23}(c) for $\varepsilon=1.2$. CS is observed between the circuits for larger value of the coupling strength and the experimental wave forms for suitable $\varepsilon$ is shown in Fig.~\ref{fig23}(b) exhibiting CS. Figure \ref{fig23}(d) shows the numerically obtained time traces of the two systems displaying CS for $\varepsilon=2.0$.

PS between the circuits can be once again visualized by plotting the localized sets. Figure \ref{fig26} demonstrates the experimentally observed attractors along with the sets. In Figs.~\ref{fig26}(a) and \ref{fig26}(b) the sets are distributed over the entire attractor for lower values of coupling strength due to the absence of PS. The corresponding numerically obtained figures are plotted in Figs.~\ref{fig26}(e) and \ref{fig26}(f) for $\varepsilon=0$. If we increase the coupling strength to a sufficiently large value, the sets are localized on the attractor as depicted in Figs. \ref{fig26}(c) and \ref{fig26}(d) confirming the phase locking of both systems. The corresponding numerically obtained figures are plotted in Figs.~\ref{fig26}(g) and \ref{fig26}(h) for $\varepsilon=1.2$.
\begin{figure}
\centering
\includegraphics[width=0.6\columnwidth]{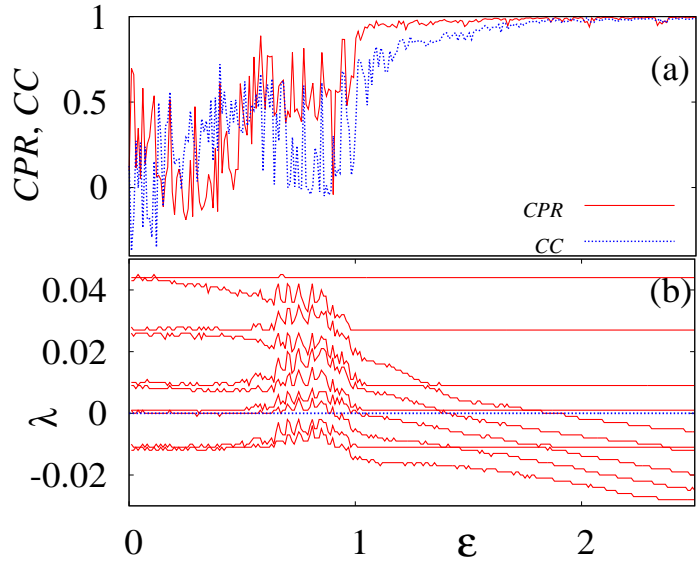}
\caption{\label{fig28} (Color online) (a) CPR (continuous line), CC (dotted line) and (b) spectrum of maximal LEs for subsystem coupling configuration with $(\beta_{1},\beta_{2})=(1,-1)$ in subsystem coupling configuration as a function of coupling strength $\varepsilon \in(0,2.5)$.}
\end{figure}

In the absence of the coupling the systems are evolving freely without any synchronization with six positive LEs (three for each systems) and so CPR and CC $\approx0$. Beyond $\varepsilon>1.0$ ($\varepsilon=1.02$), CPR becomes $\approx 0.97$ which confirms the existence of PS which is depicted in Fig.,~\ref{fig28}(a). At this value of $\varepsilon$, the zeroth LE of the coupled system becomes negative which is shown in Fig.~\ref{fig28}(b). If we increase the coupling strength further, the CC of the systems increases and reaches the unit value at $\varepsilon=1.91$ which is evident from Fig.~\ref{fig28}(a) and for this value of coupling strength all the positive LEs (except three positive LEs) become negative confirming the existence of CS in the coupled time-delay systems [Fig.~\ref{fig28}(b)].
\begin{figure}
\centering
\includegraphics[width=0.6\columnwidth]{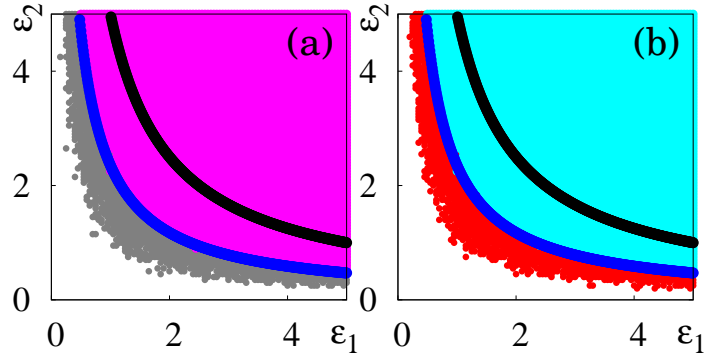}
\caption{\label{fig29} (Color online) (a), (b) Numerically obtained two parameter diagrams in ($\varepsilon_{1}-\varepsilon_{2}$) plane shows various types of synchronization states in subsystem coupling configuration for ($\beta_{1},\beta_{2})=(1,1$) and ($\beta_{1},\beta_{2})=(1,-1$), respectively. Color codes are similar to the Fig.~\ref{fig15}.}
\end{figure}

The numerically obtained phase diagrams in the $(\varepsilon_1-\varepsilon_2)$ parameter plane of coupling strengths are plotted to identify the global picture of the regimes of different types of synchronization states of the coupled time-delay systems (Eq.~\ref{eqn12}) in Fig.~\ref{fig29}(a) and \ref{fig29}(b) for ($\beta_{1}, \beta_{2}) = (1,1$) and ($1,-1$), respectively. In these figures, we mark PS and IPS regimes when CPR $>0.96$. Also CS region is marked when CC $>0.96$ and IS region is marked when CC $>-0.96$. In both figures the white color represents the desynchronized state, grey color in Fig.~\ref{fig29}(a) and red color (dark grey) in Fig.~\ref{fig29}(b) correspond to IPS and PS regimes, respectively. IS is represented by the pink color (light grey) in Fig.~\ref{fig29}(a) and the light blue (light grey) color indicate CS region in Fig.~\ref{fig29}(b). We have also carried out a linear stability analysis for the above coupling configuration as in Sec.~\ref{sec:level3b} and it gives a condition for the stability of the synchronized states as $\varepsilon_{1c}>\frac{2|k(\beta\phi-\alpha)|}{\varepsilon_{2c}}$. From this relation one can obtain the threshold value of the coupling strength for different synchronized states. For example, we choose $\phi$ values as in Sec.~\ref{sec:level3} and the feedback ($\beta_{1},\beta_{2})=(1,1)$, we yield a transition from no synchronization to an IS state which is plotted along with the numerically obtained synchronized region. The black points in Fig. \ref{fig29}(a) correspond to the outer regimes of the piecewise linear function (\ref{eqn2}) and the blue (dark grey) filled circles indicate the stability state of the middle regime. Similar to Fig.\ref{fig15}, the numerically obtained IS regime exactly fits with the analytically obtained middle regime of the piecewise linear function. Figure ~\ref{fig29}(b) shows the transition curves for the case of $(\beta_{1},\beta_{2})=(1,-1)$ which indicates the transition from no synchronization to CS in the subsystem coupling configuration.
\section{\label{sec:level5}Conclusion}
In this paper, we have experimentally demonstrated the occurrence of various types of synchronization in coupled piece-wise linear time-delayed electronic circuits with threshold nonlinear function where the circuits are coupled indirectly through a common dynamic environment. We have carried out these studies in two different coupling configurations, namely mutual and subsystem coupling configurations. In both configurations, depending upon the strength of the coupling and the nature of the feedback, we observe different types of synchronization transitions such as transition from IPS to IS and from PS to CS in hyperchaotic regimes. Snapshots of the time evolution, phase projection and localized sets plots of the circuits observed from the oscilloscope confirm the various synchronization phenomenon experimentally. The corresponding numerical simulations are also presented in detail. Further, the transition to different synchronization states can be verified from the changes in the maximal LEs, index CPR and CC of the coupled systems as a function of the coupling strength. Also we have presented a detailed linear stability analysis to obtain synchronization conditions for different synchronized states. From our investigation,  it is also clear that the concept of localized sets and recurrence quantification measures can be potentially used to analyze other experimental data and to identify the underlying synchronization transitions.

It is also to be noted that the study carried out in this paper can also be extended to other environmentally coupled systems with and without delay because the coupling scheme does not alter the system dynamics in phase-space (the structure of the systems is the same for both synchronized and desynchronized states). In particular, the dynamic coupling occurs in many biological systems (where delay is inherent in general) where the coupling occurs through many steps and direct coupling is often considered as a simplified concept to interpret the dynamics. An important example of such systems is the population of cells in which oscillatory reactions are taking place, which communicate via chemicals that diffuse in the surrounding medium \cite{katriel08,kruse05,schibler05}.

\section*{Acknowledgments}
The work of R.S., K.S. and M.L. has been supported by the Department of Science and Technology (DST), Government of India sponsored IRHPA research project. M.L. has also been supported by a DST Ramanna project and a DAE Raja Ramanna Fellowship. D.V.S. and J.K. acknowledge the support from EU under project No. 240763 PHOCUS(FP7-ICT-2009-C).

\end{document}